\documentclass[12pt]{article}
\usepackage{epsfig}
\begin{document}
\begin{center}
{\Large {\bf 2D $R^2$ gravity in weak coupling limit. }}

{
\vspace{1cm}
{M.A.~Zubkov$^a$  }\\
\vspace{.5cm}{\it $^a$ ITEP, B.Cheremushkinskaya 25, Moscow, 117259, Russia }}
\end{center}
\begin{abstract}
2D $R^2$ quantum gravity in infinitely large invariant volume is considered. In
weak coupling limit the dynamics is reduced to quantum mechanics of a single
degree of freedom. The correspondent two - point Green function is calculated
explicitly in Gaussian approximation.
\end{abstract}

\today \hspace{1ex}

\newpage

Quantum gravity is still one of the most important problems of modern
theoretical physics. The possible data on the structure of gravity at the Plank
scale comes from the only real experiment that is the history of the Universe
as a whole and might be extracted only from the astrophysical observations. The
correspondent energies can not be achieved in usual experiments, which are
"made by hands". And this complicates the issue.

Mathematical structure of quantum gravity should be related to Riemannian
geometry: It must appear in the classical limit (after the Wick rotation). Of
course, there exist a lot of different geometrical and algebraic structures,
that could serve as a basis of correct quantum gravity. However, right now we
have no reason for making a preference. Therefore, it is natural to make a
"minimal" choice: to construct the theory that is based on the Riemannian
geometry itself rather than on something that involves it in a certain limit.
Even if such a theory would be incorrect, it might appear as a good
approximation at low enough energies.

When dealing with quantization of Riemannian geometry with the simplest choice
of Einstein - Hilbert action we encounter difficulties related to its
unboundedness from below. As a result vacuum is expected to be dominated by
fractal structures with large values of scalar curvature. This guess is
justified by numerical simulations of various discretized models \cite{Loll}.
Next difficulty of this choice of the action is that the theory appears to be
nonrenormalizable \cite{nrenorm}.  All this leads to necessity to include into
the action the terms, that might made it bounded below while providing the
theory to be renormalizable. This is indeed achieved if we make the simplest
choice of the gravity with squared curvature action \cite{R2}. Besides, it can
be constructed in such a way, that Einstein equations without matter appear in
the classical limit \cite{Z2003}. In addition, the theory in this case is
asymptotic free. While direct inclusion of matter destroys the correct
classical limit, the matter can be introduced into the model as singularities
of space - time. In this case classical Einstein equations survive
\cite{Z2003,mass_singularities}. Besides, there was noticed that the unitarity
problem appeared in the second derivative theories of gravity might be resolved
nonperturbatively \cite{unitarity}. This enforces us to consider this model as
a good candidate for the realistic quantum gravity theory.

One of the most fruitful ways of nonperturbative investigations in quantum
field theory (and, probably, the only one that can be treated as exhaustive) is
the numerical lattice method\footnote{The discretizations of quantum gravity
model with the squared curvature action of general form were suggested in
\cite{Z2003,Regge_R2}.}. The mentioned above model being transferred to lattice
looses considerable amount of symmetry, which must be restored in the continuum
limit. Therefore, the comparison of lattice results with analytical ones are
necessary. However, the 4D models are quite complicated and computer
simulations are rather time consuming. That's why tests of various simplified
forms of the model are of especial importance. One of the possible ways to
perform such tests is to investigate 2D $R^2$ gravity\footnote{See also
\cite{Odintsov}, where the $2D$ dilaton gravity (which can reproduce $2D$ $R^2$
gravity) was considered.}.

When dealing with Euclidean $2D$ $R^2$ quantum gravity the question arises:
what should we measure? The often used quantity is the partition function at
fixed invariant volume $V$ as a function of this volume.
\begin{eqnarray}
Z(V) & = & \int Dg {\rm exp} ( - \beta \int R^2 \sqrt{g} d^2 x)\delta(\int
\sqrt{g}
d^2 x - V)\nonumber\\
& = & \frac{1}{V}\int Dg {\rm exp} ( - \frac{\beta}{V} \int R^2 \sqrt{g} d^2
x)\delta(\int \sqrt{g} d^2 x - 1) , \label{Z}
\end{eqnarray}
where $\beta$ is the inverse gravitational coupling.

In the limit $\beta/V >> 1$ it can be evaluated as follows. First of all,
nontoroidal topology is suppressed because torus is the only two - dimensional
surface, which can be provided with zero curvature. Next, at $\beta \rightarrow
\infty$ the system (\ref{Z}) becomes equal to the set of $\cal N$ oscillators,
where $\cal N$ is the number of degrees of freedom. For the discretized model
this number is $N \frac{D(D-1)}{2} = N$, where $N$ is the number of points of
the discretization. As a result $Z(V) \sim \frac{1}{V}[\frac{V}{\beta}]^{N/2}$.
On the other hand, if one uses the dimensional regularization
\cite{2DR2perturb} or  the conformal field theory technic \cite{R2CFT}, the
expression for the partition function is different: $Z(V) \sim \frac{1}{V}$.
That's why we guess that the quantity (\ref{Z}) (as well as any other one that
contains the dependence upon the overall volume) is not quite acceptable for
testing the discretization of quantum gravity.

In order to find the quantities that do not suffer from the ambiguity similar
to the one mentioned above, we are forced to consider local gauge invariant
variables made of the metric tensor $g$. In this Letter we consider a choice of
the full set of such variables connected with the so - called synchronic
reference frame. The dynamics written in these variables appears to be
surprisingly simple in the weak coupling limit, i.e. at distances much less
than $\sqrt{\beta}$. We also assume that the considered distances are much less
than $\sqrt{V}$, or, equivalently, the overall invariant volume is implied to
be infinite.

Let us fix a point $A$ and consider the geodesic lines ended at $A$. Close to A
those lines are seen as straight. Let us fix one of these lines ($l_0$) and
parameterize the others $l_{\sigma}$ by the (invariant) angle $\sigma$ between
$l_0$ and $l_s$. At $\beta \rightarrow \infty$ the manyfold is smooth and
$\sigma \in [0, 2\pi[$.

Let us denote by $B_{\tau}(s) \in l_{\sigma}$ such point that the distance
between $A$ and $B_{\tau}(\sigma)$ is equal to $\tau =
\rho(A,B_{\tau}(\sigma))$. These points form the closed curve ${\cal C}_{\tau}
= \{ B_{\tau}(\sigma) | \sigma \in [0, 2\pi[ \}$. The length of the piece of
the curve that connects $B_{\tau}(\sigma)$ and $B_{\tau}(0)$ is denoted by
$s(\sigma, \tau)$.

The function $s(\sigma, \tau)$ describes completely local metric properties of
the manyfold. Let us parameterize points of the manyfold by variables $(\tau,
\sigma)$. Then
$ g(\tau, \sigma) = \left(\begin{array}{cc} 1 & 0 \\
0 & e^{\phi(\tau, \sigma)}\end{array}\right),$
$s(\tau,\sigma) = \int_0^{\sigma} d \sigma_1 e^{\phi(\tau, \sigma_1)/2}$, and
\begin{equation}
R = - [\partial_{\tau}^2 \phi + \frac{1}{2} (\partial_{\tau}\phi)^2].\label{R}
\end{equation}

It is worth mentioning that 2D quantum gravity without $R^2$ term in the action
was considered in the synchronic reference frame in a number of papers (see,
for example, \cite{R2synchronic}).

The curvature (\ref{R}) depends on "time" derivatives $\partial_{\tau}$
only(and not on the "space" derivatives $\partial_{\sigma}$). This simplifies
the dynamics considerably. The partition function can be rewritten as follows:
\begin{equation}
Z = \int D\phi {\rm exp} ( - \beta \int [\partial_{\tau}^2 \phi + \frac{1}{2}
(\partial_{\tau}\phi)^2]^2 e^{\phi/2} d\sigma d\tau).\label{Z_1}
\end{equation}
(At $\beta \rightarrow \infty$ the Faddeev - Popov determinant is irrelevant.)

Minimum of the exponential factor in (\ref{Z_1}) is achieved if
\begin{equation}
\partial_{\tau}^2 \phi + \frac{1}{2} (\partial_{\tau}\phi)^2 = 0 \label{R_1}
\end{equation}
Solving (\ref{R_1}) with the appropriate boundary conditions at $\tau = 0$ we
get:
$\phi_{cl} =  2\, {\rm log} \, \tau$.
This implies, in particular, that the length of ${\cal C}_{\tau}$ is equal to
$2 \pi \tau$ as it should for the flat geometry.

Therefore we represent
\begin{equation}
\phi =  2\, {\rm log} \, \tau + \omega (\tau, \sigma)
\end{equation}
and our new dynamical variable is $\omega$, that represents the deviation from
flatness.  The Fourier series for $\omega$ is:
$\omega(\tau, \sigma) = \sum_n \omega_n(\tau) e^{i n \sigma}$, where
$\omega_{n} = \omega^+_{-n}$. We denote $\omega_n = \omega_n^{(1)} + i
\omega_n^{(2)}$ with real $\omega^{(k)}_n$. Obviously, $\omega_0^{(2)} = 0$.
Gaussian approximation gives:
\begin{equation}
Z = \int \Pi_n D\omega_n {\rm exp} ( - 2\pi \beta \sum_{n\ge 0} \int
|\partial_{\tau}^2 \omega_n + \frac{2}{\tau} \partial_{\tau}\omega_n|^2 \tau
 d\tau).\label{Z2}
\end{equation}

One can see that all harmonics propagate independently and our aim is to
consider the propagation of a single harmonic. With the boundary condition
$\omega = 0$ at $\tau = 0$ we obtain the following expression for the two -
point Green function (see Appendix):
\begin{equation}
 <\omega^{(k)}_n(\tau) \omega^{(k^{'})}_{n^{'}}(\tau^{'})> = \frac{\delta_{n n^{'}}\delta_{k k^{'}}}{24 \pi \beta}\{ \tau^2(1 - \frac{\tau}{2\tau^{'}}) \theta(\tau^{'} - \tau) +
 (\tau^{'})^2(1 -  \frac{\tau^{'}}{2\tau}) \theta(\tau - \tau^{'})
 \}\label{FIN}
\end{equation}
(for $(k,n) \ne (2,0)$).

This expression exhausts the description of weak coupling 2D $R^2$ gravity in
Gaussian approximation. One of the most interesting features of (\ref{FIN}) is
that the correlation is not absent even if one of the points is at $\tau =
\infty$.

Now we can calculate, in particular, the correlation of lengths $|{\cal
C}_{\tau}|$ of the curves ${\cal C}_{\tau}$ at different $\tau$:
\begin{equation}
 < [\frac{|{\cal C}_{\tau}|}{2\pi \tau} - 1] [\frac{|{\cal C}_{\tau^{'}}|}{2\pi \tau^{'}}-1] > = \frac{1}{96 \pi \beta}\{ \tau^2(1 - \frac{\tau}{2\tau^{'}}) \theta(\tau^{'} - \tau) +
 (\tau^{'})^2(1 -  \frac{\tau^{'}}{2\tau}) \theta(\tau - \tau^{'}) \}
\end{equation}
and the fluctuation\footnote{(\ref{C}) gives us the final estimate for the
application of the considered approximation: $\tau << \sqrt{192 \pi \beta} \sim
25 \sqrt{\beta}$ and $\tau << \sqrt{V/\pi}$.} of this length as a function of
$\tau$:
\begin{equation}
 <[\frac{|{\cal C}_{\tau}|}{2\pi \tau}-1]^2> = \frac{1}{192 \pi \beta} \tau^2\label{C}
\end{equation}

To conclude, we have defined dynamical variables that describe completely local
properties of 2D quantum gravity. In weak coupling their Fourier modes
propagate independently. The correspondent quantum mechanical problem is
reduced in Gaussian approximation to calculation of the two - point function.
The result is relatively simple and is given by (\ref{FIN}). The obtained
expressions (\ref{FIN}) - (\ref{C}) could be used, for example, for examining
various lattice discretizations of quantum gravity models.

The author is grateful to E.T.Akhmedov for useful discussions. This work was
partly supported by RFBR grants 03-02-16941 and 04-02-16079, by the INTAS grant
00-00111, the CRDF award RP1-2364-MO-02, by Federal Program of the Russian
Ministry of Industry, Science and Technology No 40.052.1.1.1112.

\section{Appendix}

 We change variables
and introduce $t = \frac{1}{\tau}$. The resulting quantum mechanical partition
function is
\begin{equation}
Z_0 = \int D\omega {\rm exp} ( - 2\pi \beta \int (\omega^{''})^2
 t^5 d t).\label{Z3}
\end{equation}

The derivative $\partial_{\tau} \omega$ is bounded at $\tau = 0$ due to
smoothness of the surface. Therefore $\omega^{'} = - \frac{1}{t^2}\partial_{
\tau} \omega$ should tend to zero at infinite t. Thus, the system (\ref{Z3}) is
complemented by the boundary conditions $\omega, \omega^{'} \rightarrow 0$ at
$t \rightarrow \infty$.

We then use the auxiliary variable $\eta = \omega^{'}$. It propagates according
to the wave function
\begin{equation}
\Psi(\eta, t) = \int D \bar{\eta} {\rm exp} [ - 2\pi \beta \int_0^{\infty}
(\bar{\eta}^{'})^2
 t^5 d t]\, \delta( \bar{\eta}(t) - \eta),\label{Psi}
\end{equation}
where $\bar{\eta} (\infty) = 0$.

We define the hamiltonian $H(t)$:
\begin{eqnarray}
 <\eta_1|e^{- H(t) \Delta t)}|\eta_2> & = & \int D \bar{\eta} {\rm exp} [ - 2\pi
\beta \int_t^{t + \Delta t} (\bar{\eta}^{'})^2
 t^5 d t]\nonumber\\ &&\delta(\bar{\eta}(t) - \eta_1)\, \delta(\bar{\eta}(t+\Delta t) - \eta_2).\label{H}
\end{eqnarray}

Standard methods give
\begin{equation}
<\eta_1|e^{- H(t) \Delta t)}|\eta_2> = {\rm exp} ( - 2\pi \beta (\eta_2 -
\eta_1)^2 t^5 /\Delta t)
\end{equation}
and
\begin{equation}
<\eta_1|e^{- \int_{t_1}^{t_2} H(t) d t}|\eta_2> = {\rm exp} ( - 8\pi \beta
(\eta_2 - \eta_1)^2 /(\frac{1}{t_1^4} - \frac{1}{t_2^4}))
\end{equation}
Therefore
\begin{eqnarray}
\Psi(\eta, t)& = & \int d \eta_1 <\eta_1|e^{- \int_{0}^{t} H(t^{'}) d
t^{'}}|\eta> <\eta|e^{- \int_{t}^{\infty} H(t^{'}) d t^{'}}|0> \nonumber\\
& = &  <\eta|e^{- \int_{t}^{\infty} H(t^{'}) d t^{'}}|0>  =  {\rm exp} ( - 8\pi
t^4 \beta \eta^2 )
\end{eqnarray}

We have the following result for the two point function:
\begin{eqnarray}
 <\eta(t) \eta(t^{'})> & = & \frac{1}{Z_0}\int d \eta d \eta^{'} \eta \eta^{'}
\nonumber \\ &&  {\rm exp} ( - 8\pi (t^{'})^4 \beta (\eta^{'})^2 - 8\pi \beta
(\eta - \eta^{'})^2 /(\frac{1}{t^4} - \frac{1}{(t^{'})^4}))
\end{eqnarray}
at $t^{'} > t$. Thus
\begin{equation}
<\eta(t) \eta(t^{'})> =  \frac{1}{16\pi \beta (t^{'})^4} \theta(t^{'} - t) +
\frac{1}{16\pi \beta t^4} \theta(t - t^{'})
\end{equation}
Let $G(t, t^{'}) = <\omega(t) \omega(t^{'})>$. Then
\begin{equation}
\partial_{t}\partial_{t^{'}} G(t, t^{'}) = \frac{1}{16\pi \beta (t^{'})^4}
\theta(t^{'} - t) + \frac{1}{16\pi \beta ^4 t} \theta(t - t^{'})
\end{equation}
and
\begin{equation}
 G(t, t^{'}) = \int_t^{\infty} d \bar{t} \int_{t^{'}}^{\infty} d \bar{t^{'}}
\frac{1}{16\pi \beta (\bar{t}^{'})^4} \theta(\bar{t}^{'} - \bar{t}) +
\frac{1}{16\pi \beta ^4 \bar{t}} \theta(\bar{t} - \bar{t}^{'})
\end{equation}

The final solution is
\begin{equation}
 G(t, t^{'}) = \frac{1}{48 \pi \beta}\{\frac{2 t -  t^{'}}{ t ^3} \theta(t - t^{'}) +
\frac{2 t^{'} -  t}{ (t^{'}) ^3} \theta(t^{'}-t) \}
\end{equation}

Coming back to variable $\tau$ we obtain:
\begin{equation}
 <\omega(\tau) \omega(\tau^{'})> = \frac{1}{24 \pi \beta}\{ \tau^2(1 - \frac{\tau}{2\tau^{'}}) \theta(\tau^{'} - \tau) +
 (\tau^{'})^2(1 -  \frac{\tau^{'}}{2\tau}) \theta(\tau - \tau^{'})
 \} \label{F}
\end{equation}

This expression can be checked as follows. For any function $f(\tau)$ vanishing
at $\tau = 0, \infty$ together with its derivatives (up to the third order)  we
have:
\begin{eqnarray}
&& \int 4\pi \beta \, f(\tau) \,(\partial_{\tau}^2 - 2 \partial_{\tau}
\frac{1}{\tau})\tau (\partial_{\tau}^2 + \frac{2}{\tau}) <\omega(\tau)
\omega(\tau^{'})>  d \tau =
\nonumber\\
&& = \int \frac{1}{6}\{ 24\, \delta(\tau - \tau^{'}) \nonumber\\
&& + [-20 \tau +25 \frac{\tau^2}{\tau^{'}} - \frac{\tau^3}{(\tau^{'})^2}
 - 4 \frac{(\tau^{'})^2}{\tau}] \delta^{'}(\tau - \tau^{'})\nonumber\\
\nonumber\\
&& + [-10 \tau^2 + 7 \frac{\tau^3}{\tau^{'}} + 2 (\tau^{'})^2 +
\frac{\tau^3}{\tau^{'}}]\delta^{''}(\tau - \tau^{'})\nonumber\\
\nonumber\\
&& + [- \tau^3 +  \frac{\tau^4}{2\tau^{'}} + \tau (\tau^{'})^2 -
\frac{\tau^3}{2}]\delta^{'''}(\tau - \tau^{'})\} f(\tau) d \tau
\end{eqnarray}

Using integration by parts this expression can be rewritten as
\begin{eqnarray}
&& \int 4\pi \beta (\partial_{\tau}^2 - 2 \partial_{\tau} \frac{1}{\tau})\tau
(\partial_{\tau}^2 + \frac{2}{\tau}) <\omega(\tau) \omega(\tau^{'})> f(\tau) d
\tau =
\nonumber\\
&& = \int  \, \delta(\tau - \tau^{'}) f(\tau) d \tau
\end{eqnarray}

This verifies that (\ref{F}) is indeed Green function of the operator $4 \pi
\beta (\partial_{\tau}^2 - 2 \partial_{\tau} \frac{1}{\tau})\tau
(\partial_{\tau}^2 + \frac{2}{\tau})$ encountered in (\ref{Z2}).


\begin{thebibliography}{99}


\bibitem{Loll}
R. Loll, gr-qc/9805049; Living Reviews in Relativity, www.livingreviews.org

\bibitem{nrenorm}
M.H.Goroff, A.Sagnotti, Nucl. Phys. {\bf B 266}, 709 (1986)

\bibitem{R2}
K.S.Stelle, Phys. Rev. {\bf D 16} , 953 (1977)

I. G. Avramidi, Soviet Journal of Nuclear Physics, 44 (1986) 160-164,  "Heat
Kernel and Quantum Gravity", Lecture Notes in Physics, Series Monographs, LNP:
m64 (Berlin: Springer-Verlag,  2000), hep-th/9510140

\bibitem{Z2003}
M.A.Zubkov, Phys. Lett {\bf B 582}, 243 (2004)

\bibitem{mass_singularities}
L.Infeld, Rev. Mod. Phys. {\bf 29}, 398 (1957)

\bibitem{unitarity}
E.T.Tomboulis, Phys. Rev. Lett. {\bf 52}, 1173 (1984)


\bibitem{Regge_R2}
Herbert W. Hamber, Ruth M. Williams, Nucl.Phys. {\bf B 248} (1984) 392,
Nucl.Phys. {\bf B 269} (1986) 712, Nucl.Phys. {\bf B435} (1995) 361

\bibitem{Odintsov}
E.Elizalde, S. Naftulin and S.D. Odintsov, Phys. Lett. {\bf B 323}, 124 (1994)

S.Naftulin and S.D. Odintsov, Mod.Phys.Lett. {\bf A 10}, 2071 (1995)


\bibitem{2DR2perturb}
J.Nishimura, S.Tamura, A.Tsuchiya, Mod.Phys.Lett. {\bf A9} 3565 (1994)


\bibitem{R2CFT}
H.Kawai, R.Nakayama, Phys. Lett. {\bf B306}, 224 (1993)

\bibitem{R2synchronic}
H.Kawai, N.Kawamoto, T.Mogami, Y.Watabiki, Phys. Lett. {\bf B306}, 19 (1993)





\end{thebibliography}
\end{document}